\documentclass[useAMS,usenatbib,referee]{biom2} 

 \newcommand{\bitem}{\begin{itemize}}
 \newcommand{\eitem}{\end{itemize}}

\usepackage[figuresright]{rotating}
\usepackage{graphicx}
\usepackage{amsmath}
\usepackage{bm}
\usepackage[shortlabels]{enumitem}
\usepackage{amssymb}
\usepackage{natbib}
\usepackage{multirow}
\usepackage{threeparttable}
\usepackage{enumitem}

\usepackage[colorinlistoftodos]{todonotes} 

\title[]{Change-point Detection for Piecewise Exponential Models}

\author
{Philip Cooney\emailx{phcooney@tcd.ie} \\
School of Computer Science and Statistics,  Trinity College Dublin, Dublin, Ireland
\and
Arthur White\emailx{arwhite@tcd.ie} \\
School of Computer Science and Statistics,  Trinity College Dublin, Dublin, Ireland}

\begin{document}



\date{September 2021.}
\pagerange{\pageref{firstpage}--\pageref{lastpage}} 
\volume{}
\pubyear{}
\artmonth{}

\doi{}

\label{firstpage}


\begin{abstract}
\textbf{Background}. In decision modelling with time to event data, parametric models are often used to extrapolate the survivor function. One such model is the piecewise exponential model whereby the hazard function is partitioned into segments, with the hazard constant within the segment and independent between segments and the boundaries of these segments are known as change-points. \textbf{Objective}. We present an approach for determining the location and number of change-points in piecewise exponential models. \textbf{Methods}. Inference is performed in a Bayesian framework using Markov Chain Monte Carlo (MCMC) where the model parameters can be integrated out of the model and the number of change-points can be sampled as part of the MCMC scheme. We can estimate both the uncertainty in the change-point locations and hazards for a given change-point model and obtain a probabilistic interpretation for the number of change-points. \textbf{Results}. We evaluate model performance to determine changepoint numbers and locations in a simulation study and show the utility of the method using two data sets for time to event data. In a dataset of Glioblastoma patients we use the piecewise exponential model to describe the general trends in the hazard function. In a data set of heart transplant patients, we show the piecewise exponential model produces the best statistical fit and extrapolation amongst other standard parametric models. \textbf{Conclusions}. Piecewise exponential models may be useful for survival extrapolation if a long-term constant hazard trend is clinically plausible. A key advantage of this method is that the number and change-point locations are automatically estimated rather than specified by the analyst.

\end{abstract}



\maketitle

\section*{Highlights}
\begin{itemize}
  \item Presentation of a novel method to identify the location and number of change-points in a piecewise exponential model.
  \item Method implemented within a freely available R package which produces outputs relevant for health economists conducting survival analysis.
  \item Particularly useful for identifying the final change-point after which the \underline{observed} hazards are approximately constant.
  \item Addresses some of the key concerns raised in NICE TSD 21 regarding piecewise exponential models.
\end{itemize}

\section{Introduction \label{section intro}}

In survival analysis, one is interested in the lifetime of an individual and its hazard rate function. The hazard rate quantifies the instantaneous failure rate of a subject who has not failed at a given time point. Because the survival probabilities are directly related to the integral of the hazard function, changes in this function over time are of interest in a variety of situations. Parametric survival models allow the hazard function to vary over time (e.g. monotonic change associated with the Weibull distribution), however, there are examples when there appears to be a sudden change in the hazard. In these situations, piecewise exponential models, which allow the hazard to change at distinct time points, but constrain the hazard to be constant within each interval can provide a better fit to the observed data and allow for straightforward interpretation of the hazard function. 

\cite{Davies.2013} discuss some of the issues regarding extrapolating treatment benefits for new technologies based on limited clinical trial data while \cite{Kearns.2019} highlight that standard parametric models may not be flexible enough to accurately fit the observed data. One approach to extrapolating long term survival is noted by \cite{Bagust.2014} who suggest identifying whether there is evidence of long-term linear trends in each arm in the latter part of the data, when transient effects have dissipated. They suggested assessing the cumulative hazard plots, whereby a linear trend indicates a constant hazard, however, because visual inspection is subjective a statistical analysis is more appropriate. 

Following \cite{Matthews.1982}, the problem of estimating a single change-point using Maximum Likelihood (ML) estimation, when the hazards are unknown, was extensively studied. \cite{Yao.1986} showed that the likelihood is maximised at event times and that a change-point obtained by ML estimation is consistent i.e. it approaches the true change-point value as the sample size increases. \cite{Goodman.2011} used a Wald type statistic with an alpha spending function to preserve Type 1 error when considering multiple change-point models while \cite{Han.2014} used a likelihood ratio test with backward elimination. Comprehensive reviews of the technical aspects of change-point analysis of hazard functions are provided by \cite{Anis.2009} and \cite{Muller.1994}.

A key issue for the (frequentist) approaches described above is that the uncertainty in the change-point(s) and hazards are not assessed, which is a key feature of interest in decision making. While \cite{Loader.1991} proposed a likelihood ratio method to find confidence regions for the change-point, this has yet to be extended to multiple change-points. Additionally, frequentist procedures rely on null distributions which often require some technical assumptions and conditions that are difficult to verify in practice and may not hold for for small-to-moderate sample sizes. Even in the presence of larger sample sizes, likelihood ratio tests will favour more complex models even when a simpler model fits the data adequately \citep{Raftery.1986}.

In contrast to frequentist methods, Bayesian approaches do not require asymptotics, instead using a set of prior beliefs which are updated using information from an observed sample. Bayesian approaches can readily characterize the uncertainty associated with the hazards and the location of change-points. \cite{Arjas.1994} introduced a Bayesian approach for multiple change-point estimation using Gibbs sampling. \cite{Kim.2020} use a stochastic approximation Monte Carlo algorithm to identify which particular number and location of change-points gives the highest log-posterior values. They allow the sampler to move between different change-point models as part of the estimation procedure, however, they do not present the relative probabilities of models with different numbers of change-points. \cite{Chapple.2020} estimate piecewise exponential (and piecewise log-linear) models using reversible jump MCMC methods \citep{Green.1995}. 

 In this paper we introduce a novel method for the estimation of piecewise exponential models with multiple change-points. We apply a reversible jump algorithm to a collapsed change-point model \citep{Wyse.2010}. Collapsing the model leads to highly efficient MCMC performance and accurate estimates of the number of and location of change-points. Estimation of the hazard rates of individual change-point segments or the overall behaviour of the hazard function is straightforward to compute using a post-hoc routine.  

The rest of this article is organized as follows. In Section \ref{section:Con-Haz-Models} we discuss the exponential and piecewise exponential models, the latter of which allows for changes in the hazard function. We also describe the approach to estimate the number of changes in the hazard function of a piecewise exponential model. Section \ref{section:Simstudy} presents a simulation study to determine the sample sizes and changes in hazards required for the adequate estimation of the change-point location and frequency. In Section \ref{section:RWE} we apply both methods to two real data sets that record  time to death of Glioblastoma patients and time to death of heart transplant patients. The former data set has only recently become publicly available and to our knowledge has not been analysed using change-point models before. A discussion of the methods concludes the paper in Section \ref{section Discussion}. The proposed method is implemented in a R package which can be installed via GitHub (link in Supporting Information).  

\section{Constant Hazard and Change-point Survival Models}
\label{section:Con-Haz-Models}

In this section, exponential and piecewise exponential models are discussed. The rationale for the manipulation of time to event data to time between events is also explained. We then present the model which estimates the number and location of the change-points. 
\subsection{Exponential model}
\label{section:Exponential}

The simplest possible survival distribution is obtained by assuming a constant risk over time, so the hazard is $\lambda(t) = \lambda$ for all $t$. The corresponding survival function is $S(t) = \exp\{-\lambda t\}$ and is known as the exponential distribution. The density function is then $f(t) = \lambda(t) S(t) = \lambda \exp\{- \lambda t\}$. Consider a sample of $n$ observations of survival times $t_{1:n} = (t_1, \dots, t_n)$ being time ordered, some of which may be censored. The likelihood function may be written as $$\pi(t_{1:n}|\lambda) = \prod_{i=1}^n \lambda^{v_i} S(t_i),$$ with $v_i = 1$ if the subject failed and 0 if censored.  


\noindent
Taking the natural logarithms, and noting that $\log S(t)$ is equal to the negative cumulative hazard function $\Lambda (t)$, we obtain the log-likelihood function

$$\log\pi(t_{1:n}|\lambda) = \sum_{i=1}^n v_i \log \lambda - \Lambda (t_i).$$

\noindent
The cumulative hazard is $\Lambda(t_i) = \lambda t_i$. Letting $D = \sum_{i=1}^n v_i$ denote the total number of observed deaths, and $ T = \sum_{i=1}^n t_i$ denote the total observation (or exposure) time, we can rewrite the log-likelihood as a function of these totals to obtain $\log\pi(t_{1:n}|\lambda) = D \log \lambda - \lambda T,$ and $\pi(t_{1:n}|\lambda) = \lambda^D  \exp^{-\lambda T}.$

\noindent
This distribution plays a central role in survival analysis, although it is commonly too simple to be useful in applications in its own right. Therefore, an extension to the exponential model which allows the hazard to change at various intervals called a piecewise exponential model is discussed in the subsequent section.  

\subsection{Piecewise Exponential Model }

 A change-point occurs at observation $q$ if $t_1, \dots , t_q$ are generated differently to $t_{q+1}, \dots , t_n$. In a piecewise constant model with one change-point, this requires that the segments $t_{1:q}$ and $t_{q+1:n}$ have a constant hazard within the segment, but independent hazards between segments. It is assumed that the change-points occur at a particular event time (and not a censoring time). Multiple change-points at specific event times can be denoted as a vector $\tau_{1:k}$, with these $k$ change-points splitting the data into $k + 1$ segments. The likelihood of the piecewise exponential model can be formulated as follows
\begin{equation}
\pi(t_{1:n}|\tau_{1:k},\lambda_{1:k+1}) =  \prod_{i=1}^{n}\Bigg\{ \prod_{j=1}^{k+1} \lambda_j^{\delta_{ij}v_i} \exp\bigg\{-\delta_{ij} \bigg[\lambda_j (t_i - \tau_{j-1}) + \sum_{g=1}^{j-1} \lambda_g(\tau_g - \tau_{g-1}) \bigg] \bigg\}\Bigg\}
    \label{eq:like-piecewise-exp}
\end{equation}

\noindent
with $v_i = 1$ if the subject was observed to fail and 0 otherwise and where $\delta_{ij} = 1$ if $t_i \in (\tau_{j-1}, \tau_{j}],$ and 0 otherwise. 

\noindent
By omitting the potential for covariates and restricting ourselves to discrete change-points, it should be noted that there is no loss of information in recasting the time ordered data as times between individual event times. We let $d$ be the number of event times and $n-d$ right censored survival times. For notational ease, we assume here that only one individual dies at each time, so that there are no ties in the data, however, the model implementation allows for tied events. Denote the ordered distinct survival times by $x_1,x_2, \dots, x_d$, so that $x_i$ is the $i^{th}$ ordered survival time. The set of individuals who are at risk at time $x_i$ will be denoted by $\mathcal{R}_i$ (the risk-set), so that  $\mathcal{R}_i$ is the set of individuals who are event-free and uncensored at a time just prior to $x_i$. We define $y_i$ as the total (sample) time between events $i$ and $i-1$ as
$$y_i = (x_i - x_{i-1}) \times \mathcal{R}_i +\sum_{j = 1}^{n}I(v_j = 0,x_{i-1}<t_j<x_i)\times(x_i-t_i).$$
This is the composed of the difference between event times multiplied by the risk set at the event time plus the difference between any censored observations and the previous event time $x_{i-1}$, provided they occurred within the interval $(x_{i-1},x_i)$. 

\noindent
We can re-express the likelihood of the piecewise exponential model in terms of $y_{1:d}$. Let $s_{1:k}$ be a vector representing the number of events which have occurred at each of the elements of $\tau_{1:k}$, with $s_0 = 0$ and $s_{k+1} = d$. The likelihood of interval $j$ is $\lambda_j^{s_j-s_{j-1}} \exp \left \lbrace{-\lambda_j\sum_{i=s_{j-1}+1}^{s_j} y_i} \right \rbrace$. Censored observations are also allowed, providing exposure time within intervals without an event. The likelihood is then
\begin{equation*}
    \pi(y_{1:d}|s_{1:k},\lambda_{1:k+1}) = \prod_{j=1}^{k+1} \Bigg[ \lambda_j^{s_j - s_{j-1}}  \exp\left \lbrace{-\lambda_j\sum_{i=s_{j-1}+1}^{s_j} y_i} \right\rbrace\Bigg].
\end{equation*}

\subsection{Estimation using Collapsing Change-point Approach}

Markov chain samplers that jump between models with different numbers of change-points allow us to estimate posterior probabilities for candidate models while also estimating the location of change-points within each model. Introducing priors for the change-point numbers, change-point locations and hazards $\pi(k| \xi),\pi(s_{1:k}|k),\pi(\lambda_{1:k+1}|\alpha, \beta, k)$ respectively, means that we can treat the number of change-points $k$ as a random quantity to be inferred. The model posterior then becomes 
\begin{equation}
\begin{aligned}
 \pi(k,s_{1:k}, \lambda_{1:k+1}|y_{1:d}, \alpha,\beta,\xi) \propto  \pi(y_{1:d}|s_{1:k},\lambda_{1:k+1}) \pi(s_{1:k}|k)\pi(\lambda_{1:k+1}|\alpha, \beta, k)\pi(k|\xi). 
\end{aligned}   
 \label{eq:full_posterior}
 \end{equation}

\noindent
Following the approach outlined by \cite{Wyse.2010}, if we regard the hazards $\lambda_{1:k+1}$ as nuisance parameters, the posterior density of the change-point number and their respective locations is proportional to
\begin{equation*}
\pi(k,s_{1:k}|y,\alpha,\beta,\xi) \propto  \prod_{j=1}^{k+1} \pi(y_{s_{j - 1} +1:s_j}|s_{1:k}, \alpha, \beta) \pi(s_{1:k}|k)\pi(k|\xi)
\end{equation*}
where $\pi(y_{s_{j - 1} +1:s_j} | s_{1:k}, \alpha, \beta)$ denotes the marginal likelihood of the $j^{th}$ data segment. Adopting a common, independent Gamma prior $\lambda_j \sim \mathcal G(\alpha, \beta)$ for $j = 1, \ldots, k+1$ makes this quantity straightforward to calculate; see the Appendix for full details.

\noindent
Because the marginal likelihood of each data segment is available in closed form, a switch from $k$ to $k+1$ change-points, or vice-versa, does not require the design of a bijective function between support subspaces. Changes to the change-point number are proposed and accepted with Metropolis-Hastings probability $\min(1,A)$ where
\begin{equation}
A = \frac{\pi(k+1,s^{\prime}_{1:k+1}|y,\alpha, \beta, \xi)}{\pi(k,s_{1:k}|y,\alpha, \beta,\xi)}\times \frac{P(k+1,k)}{P(k,k+1)}.
\label{eq:A_eq}
\end{equation}

\noindent
The ratio of the marginal likelihoods is straightforward to compute and can be expressed as
\begin{equation}
\frac{\pi(k+1,s^{\prime}_{1:k+1}|y,\alpha, \beta,\xi)}{\pi(k,s_{1:k}|y, \alpha,\beta,\xi)} = \frac{\pi(k+1|\xi)}{\pi(k|\xi)} \frac{\pi(s^{\prime}_{1:k+1}|k+1)}{\pi(s_{1:k}|k)}  \frac{\pi(y_{s_{j - 1} +1:s_j^{'}}|\alpha, \beta)\pi(y_{s_j^{'} +1:s_{j+1}^{'}}|\alpha, \beta)}{\pi(y_{s_{j - 1} +1:s_j}|\alpha, \beta)},
\label{eq:ratio_marg_lik}
\end{equation}
where the location of the additional change-point is denoted by $s^{'}_j$. 
%
When adding a change-point in the proposal step, one of $d - k - 1$ points where there could be a change-point are randomly selected. If this point occurs in segment $j$, segments $y_{s_{j - 1} +1:s_j^{'}}$ and  $y_{s_j^{'} +1:s_{j+1}^{'}}$ are obtained, from which we calculate the marginal likelihoods and prior densities in Equation \ref{eq:ratio_marg_lik}. 
When deleting a change-point, one of the $k$ change-points are randomly selected and $y_{s_{j - 1} +1:s_j}$ becomes the new data segment where $s_j = s_{j+1}$ before deletion. 

\noindent
The probability of adding a change-point for a model with $k$ change-points is $a_k$, and $r_{k+1}$ is the probability of removing a change-point for a model with $k+1$ change-points. Clearly $r_k = 1 - a_k$, with $r_0 = 0$ and $a_{K} = 0$, for $K$ the largest change-point number under consideration, with $r_k = a_k$ for the other change-point numbers. The proposal one step transition probabilities for the number of change-points are $P(k,k+1) = \frac{a_k}{d-k-1}$ and $P(k+1,k) = \frac{r_{k+1}}{k+1}$.


\noindent 
Following the change-point number proposal step, a single change-point location is also sampled at each iteration. One of the $k$ change-points is randomly selected, and its location sampled with probability
\begin{equation*}
\pi(s_j|y,s_{j-1},s_{j+1},\alpha, \beta,k) \propto \pi(y_{s_{j-1}+1:s_j} | s_{j-1}, s_j, \alpha, \beta)\pi(y_{s_{j}+1:s_{j+1}} | s_j, s_{j+1}, \alpha, \beta)\pi(s_{1:k}|k), 
\end{equation*}
for $s_j = s_{j-1}+1, \ldots, s_{j+1}-1.$

\noindent 
Regarding priors we assign a $\text{Poisson}(\xi)$ for the number of change-points $k$. In the examples that follow, we set $\xi =1$. The prior for the change-point locations is as in \cite{Fearnhead.2006} with $n$ referring to the number of events $\pi(s_{1:k}|k)  = {n-1 \choose 2k +1}^{-1} \prod_{j=0}^k(s_{j+1}-s_j - 1).$ For the prior for each hazard, $\pi(\lambda_j|\alpha, \beta),$  we set $\alpha = 1$, and $\beta = 1$ in the case that the timescale was in years, and $\beta = 365 \text{ or } 12$ for timescales in days or months respectively. 

Although we integrate out the hazard parameters $\lambda$ from the model during this estimation scheme, it is possible to estimate the hazards for a given change-point model using the already sampled change-point locations by simulating draws from the conditional distribution $\pi(\lambda_j | y_{1:d}, s_j, s_{j-1}, \alpha, \beta),$ for $j=1, \ldots, k+1$. In effect, this introduces an extra sampling step, in which the hazards $\lambda_{1:k+1}$ are ``uncollapsed'' and sampled at each iteration, before once again being collapsed before the change-point number and locations are sampled, albeit this is done in a post-hoc fashion. The conditional distribution $\pi(\lambda_j | y_{1:d}, s_j, s_{j-1}, \alpha, \beta),$ has a gamma distribution $\mathcal G(\alpha_j^\prime, \beta_j^\prime),$ with shape $\alpha_j^\prime = s_j - s_{j+1} + \alpha$ 
and rate $\beta_j^\prime = \sum^{s_j}_{i = s_{j_1}+1} y_i + \beta$. 








\section{Simulation Study}
\label{section:Simstudy}

We conducted a simulation study to investigate the accuracy with which the model estimated the model hazards, identified the locations of the change-points, and selected the correct number of change-points. We simulated data from models with $k = 0, 1, 2$ change-points. For each model we varied the sample size and the characteristics of the hazard function. Data from each scenario was simulated 500 times. 

For the no change-point model we considered different amounts of censoring from 0\% to 50\%. For one change-point models we simulated datasets with increasing and decreasing hazards, varying the difference in the hazard between intervals along with with sample size with a change-point at 0.5, while for two change-point models we also considered bathtub and inverted bathtub hazards with change-points at 0.5 and 1. 



We assumed a study follow up of 2 years and observations with a survival time greater than this were censored. For some simulation studies we assessed the impact of censoring within the study. The censoring percentages refer to the expected proportion of events within the study follow up which are censorsed. If a censoring percentage of $50\%$ was required, censoring and event times were generated for $100\%$ (i.e. double the required percentage) of the dataset with the censorsed time following the same piecewise distribution as the event times. This ensured that the censoring of the events occurred with approximately equal probability throughout the study follow up.



 The technique presented by \cite{Castelloe.2002} was used to determine the appropriate chain length. The Potential Scale Reduction Factor (PSRF) remained below 1.02 after around 100 iterations of the model, suggesting adequate mixing beyond this time. To ensure convergence model was run for 20,750 iterations with the first 750 discarded for each of the simulation studies detailed below.  

\subsection{Simulation Study Results}

We tested the proposed method's ability to detect the absence of a change-point. We calculated number of times our models chose the (correct) null model with no change-points for 500 simulated data sets of a particular sample size ($n\in\lbrace 100, 200\rbrace$), hazard ($\lambda \in\lbrace 0.25, 0.5, 0.75 \rbrace$) and degree of censoring (0\% or 50\%). Table \ref{tab:no-change-point-collapsing} shows that the collapsing model selects the null model approximately 95\% of the time, irrespective of the sample size hazard or censoring. 



\begin{table}
\centering
\caption{Power test for the no change-point model}
 \begin{tabular}{c c c c} 
  \hline
True hazard  & Probability Correct (\%) & $n$ & Censoring (\%)  \\ 
  
\hline
\multirow{3}{*}{0.25}   &  97 & 100   & 0    \\ 
                        &  96 & 200   & 50   \\ 
                        &  96 & 100   & 50   \\ \cline{1-4}
\multirow{3}{*}{0.5}    &  95 & 100   & 0    \\ 
                        &  96 & 200   & 50   \\ 
                        &  95 & 100   & 50   \\ \cline{1-4}
\multirow{3}{*}{0.75}   &  95 & 100   & 0   \\ 
                        &  93 & 200   & 50   \\ 
                        &  96 & 100   & 50   \\ 
 \hline
\end{tabular}
\label{tab:no-change-point-collapsing}
\end{table}

 Results for one and two change-point models are reported in Table \ref{tab:simstudy-Collapsing}. These report the frequency that the correct number of change-points were identified, and the average values of $\tau$, the posterior mean of the change-point(s) (associated standard error in parentheses) for the change-point model when the correct change-point model was selected. Also reported are $\lambda$, the simulated hazards for each interval. For clarity of exposition we omit the expected posterior mean of the hazards and its standard error, noting that the accuracy of hazard estimation is determined by the accuracy of the change-point locations. Piecewise exponential times for the event times were sampled using the rpwexp function from the R package \texttt{hesim} by \cite{hesim.2020}.

For the one and two change-point simulations studies, large sample sizes and/or large changes in hazards resulted in the correct model being selected with a high probability. When changes in hazards are relatively large, the correct model is selected with high probability at all samples, while for smaller changes moderate to large samples are required. Similarly, $\tau$ is closer to the the true values of the change-point(s) and has a smaller standard error when there are large differences between the hazards and/or large sample sizes. 



\begin{table}
\caption{Simulation Study - Simulation results for Collapsing Change-point Model}
\begin{tabular}{|c|c|c|c|c|c|}
\hline
  Model                        & Parameters & $n$ = 300     & $n$ = 500     & $n$ =1000     &  $\lambda$  \\ \hline
 \multirow{2}{*}{Increasing Small}   & $\tau$        & 0.60 (0.15) & 0.56 (0.12) & 0.52 (0.07) & 0.5,0.75               \\  
                               & \% Correct  & 53 & 77 & 97       &    \\ \hline 
 \multirow{2}{*}{Increasing Large} & $\tau$ & 0.52 (0.04) & 0.51 (0.02) & 0.5 (0.01) & 0.25,0.75  \\ \ 
                           & \% Correct & 91 & 96 & 97 &  \\ \hline 
 \multirow{2}{*}{Decreasing Small} & $\tau$ & 0.52 (0.14) & 0.52 (0.11) & 0.51 (0.08) & 0.75,0.5 \\ 
                            & \% Correct & 64 & 82 & 96 &    \\ \hline 
 \multirow{2}{*}{Decreasing Large} & $\tau$ & 0.49 (0.05) & 0.49 (0.03) & 0.5 (0.01) & 0.75,0.25 \\ 
                             & \% Correct & 94 & 96 & 98 &  \\ \hline
                    
 \multirow{3}{*}{Increasing}       & $\tau_1$ & 0.56 (0.1) & 0.52 (0.06) & 0.51 (0.03) & 0.25,0.5,0.75 \\  
                             & $\tau_2$ & 1.19 (0.14) & 1.11 (0.13) & 1.03 (0.08) &     \\ 
                             & \% Correct & 27 & 59 & 93 &    \\ \hline
\multirow{3}{*}{Decreasing}       & $\tau_1$ & 0.34 (0.1) & 0.42 (0.09) & 0.47 (0.06) & 0.75,0.5,0.25\\ 
                            & $\tau_2$ & 0.96 (0.1) & 1.01 (0.09) & 1 (0.05) &    \\ 
                             & \% Correct & 19 & 48 & 90 &    \\ \hline
\multirow{3}{*}{Bathtub}       & $\tau_1$ & 0.48 (0.03) & 0.49 (0.02) & 0.5 (0.01) & 0.75,0.2,0.75 \\  
                             & $\tau_2$ & 1.02 (0.04) & 1.01 (0.02) & 1 (0.01) &    \\ 
                             & \% Correct & 94 & 94 & 96 &    \\ \hline
 \multirow{3}{*}{Invert Bathtub}       & $\tau_1$ & 0.51 (0.03) & 0.5 (0.01) & 0.5 (0.01) & 0.2,0.75,0.2 \\ 
                             & $\tau_2$ & 0.99 (0.03) & 0.99 (0.03) & 1 (0.01) &   \\ 
                             & \% Correct & 92 & 93 & 98 &    \\ \hline                          
\end{tabular}
\label{tab:simstudy-Collapsing}
\end{table}

\section{Applications \label{section:RWE}}
We applied our methods to real data sets.  In our first application, we investigate how the method can be used to explore the behaviour of the hazard. In our second example we assess the performance of the change-point model in comparison with several popular survival models in the context of survival extrapolation.

\subsection{Glioblastoma data: Identifying trends in the hazard function}

One potential application of hazard change-point analysis is the visualization of the hazard function itself. For each simulation, we ``uncollapse'' the hazards and conditional on the change-points, plot both the posterior draws and quantiles of the hazard function. We compare the hazard function estimated from the Gibbs sampler with approaches documented by \cite{Hagar.2015}, who review a variety of packages used to  estimate the hazard in time to event data for the statistical software R.  

We consider data relating to survival times for Glioblastoma, a central nervous system cancer in which prognosis remains poor. The data is available using the R package \texttt{RTCGA} developed by \cite{RTCGA} and contains a sample of 595 patients of which 446 experience death. The median survival is 1 year with approximately $5\%$ of patients surviving after 5 years.

\begin{figure}
\centering
\includegraphics[scale=0.8]{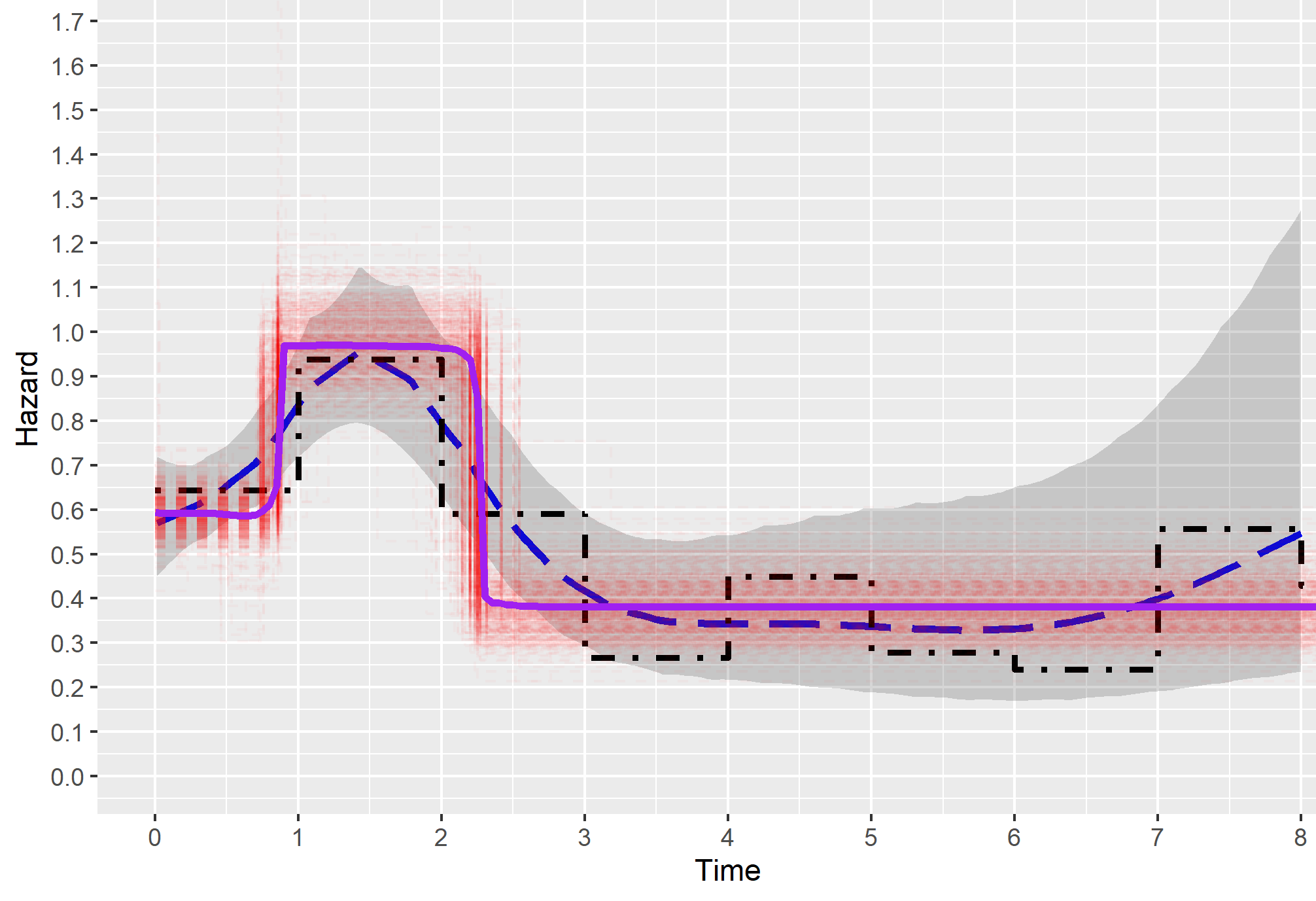}
\caption{Step (black-twodash), Smoothed (blue-longdash) and posterior mean (purple-solid) hazard functions applied to Glioblastoma data. This figure appears in color in the electronic version of this article, and any mention of color refers to that version.}
\label{fig:gioblastoma_haz}
\end{figure}

Figure \ref{fig:gioblastoma_haz} below provides an estimate of the hazard of death for the Glioblastoma data using three approaches. The first approach, coloured in black (twodash line), divides the time interval into bins of equal width (in this case one year intervals), and then estimates the hazard in each bin as the number of events $d_i$ in that bin divided by the number of patients at risk in each interval, $n_i$ with the hazard for that interval $h_i = \frac{d_i}{n_i}$ (see R package \texttt{muhaz} by \cite{muhaz.2019}). The second approach uses B-splines from a generalized linear model perspective to estimate a smoothed hazard function along with confidence regions (see R package \texttt{bshazard} by \cite{bshazard.2018}), coloured in blue (longdash line) with confidence regions in grey. The third approach, plots the posterior median of the hazard function. This figure appears in color in the electronic version of this article, and any mention of color refers to that version. 

 From Figure \ref{fig:gioblastoma_haz} it appears that the hazard peaks at 1-2 years and then falls thereafter. The posterior distribution of the change-points are concentrated at times 0.85 and 2.25 and their 95\% credible intervals do not overlap. The posterior distributions of the hazards also do not overlap with the posterior distribution of an adjacent interval, suggesting a clear change in the hazards between each interval. Using the posterior medians of the parameters we can surmise that there are three distinct intervals; in the first interval up to approximately $0.85$ years, there is a moderately large hazard of approximately $0.6$. Then from the period $0.85$ to $2.2$ years the hazard peaks around $1$ and falls to approximately $0.4$ thereafter. This finding is consistent with the other methods estimating the hazard, but which do not explicitly identify a region of peaked hazard. The finding that the hazard is peaked is consistent with \cite{Wang.2015}, however, the long term drop in hazards is more pronounced for patients in this dataset.

\subsection{Predicting survival by extrapolating constant hazards}

 \cite{Miller.1982} presented survival times for 184 patients who received heart transplants. Visual inspection of the cumulative hazard plot suggests that after 1 year the hazards are approximately constant (i.e. linear). Assuming this to be correct, we artificially censored the data at 2 years and fit the piecewise exponential model and other commonly used survival distributions to this data (using the \texttt{JAGS} program by \cite{Plummer.2003}).  For each model we assessed the statistical fit to the partially observed data and the difference in predicted survival to the fully observed data.

\begin{figure}[h]
\centering
\scalebox{1}{\includegraphics[scale=0.9]{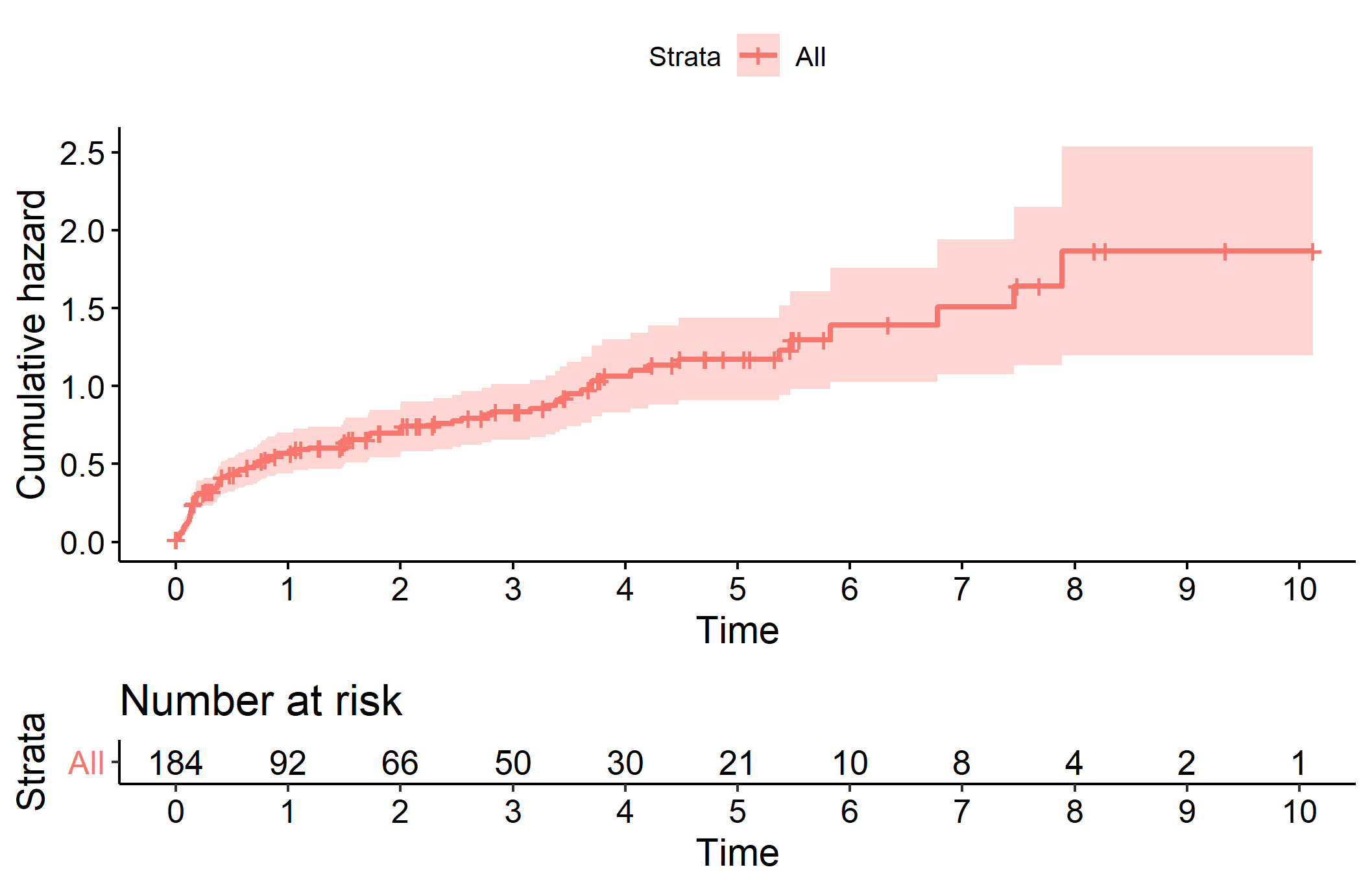} }
\caption{Cumulative Hazard Plot for Stanford Heart Data}
\label{fig:surv-gibbs2}
\end{figure}


Statistical fit was assessed through two measures, Pseudo-Marginal Likelihood (PML) and Widely Applicable Information Criterion (WAIC) with details on their respective computation available in \cite{Gelfand.1994} and \cite{Watanabe.2010}. 

Similar to the previous section we ``uncollapse'' the hazards at each simulation and calculate the survival function with the mean posterior survival being the average of these survival probabilities at each timepoint. For the piecewise wise exponential model we found that the 2 change-point model had the highest posterior probability $\approx 66\%$. The posterior mean of the first changepoint is 0.18 years at which the hazard falls from a posterior mean of 1.56 to 0.42. The posterior mean of the second change-point was 0.81 years after which the posterior mean of the hazard was 0.16.  Figure \ref{fig:surv-gibbs2} highlights that both the piecewise exponential and Weibull provide good estimates of the long term survival (with the generalized gamma reducing to the Weibull distribution), however, the Weibull distribution does not fit as well to the early part of the trial. Table \ref{tab:output-collapsing} presents the PML and WAIC (evaluated using the R package \texttt{loo} by \cite{loo.2020}) for each of the models fit to the partially observed data. Also presented is the Area Under the Curve (AUC) based on the extrapolated survival for each model along with the cumulative sum of the absolute difference (all in years) between fully observed Kaplan Meier (KM) curve and model (Absolute Difference). Consistent with the hypothesis that the long term hazards were approximately constant, the piecewise exponential approach is the best fit to the true data, both in terms of statistical fit and deviation in terms of AUC.

\begin{figure}[h]
\centering
\scalebox{0.75}{\includegraphics[scale=1]{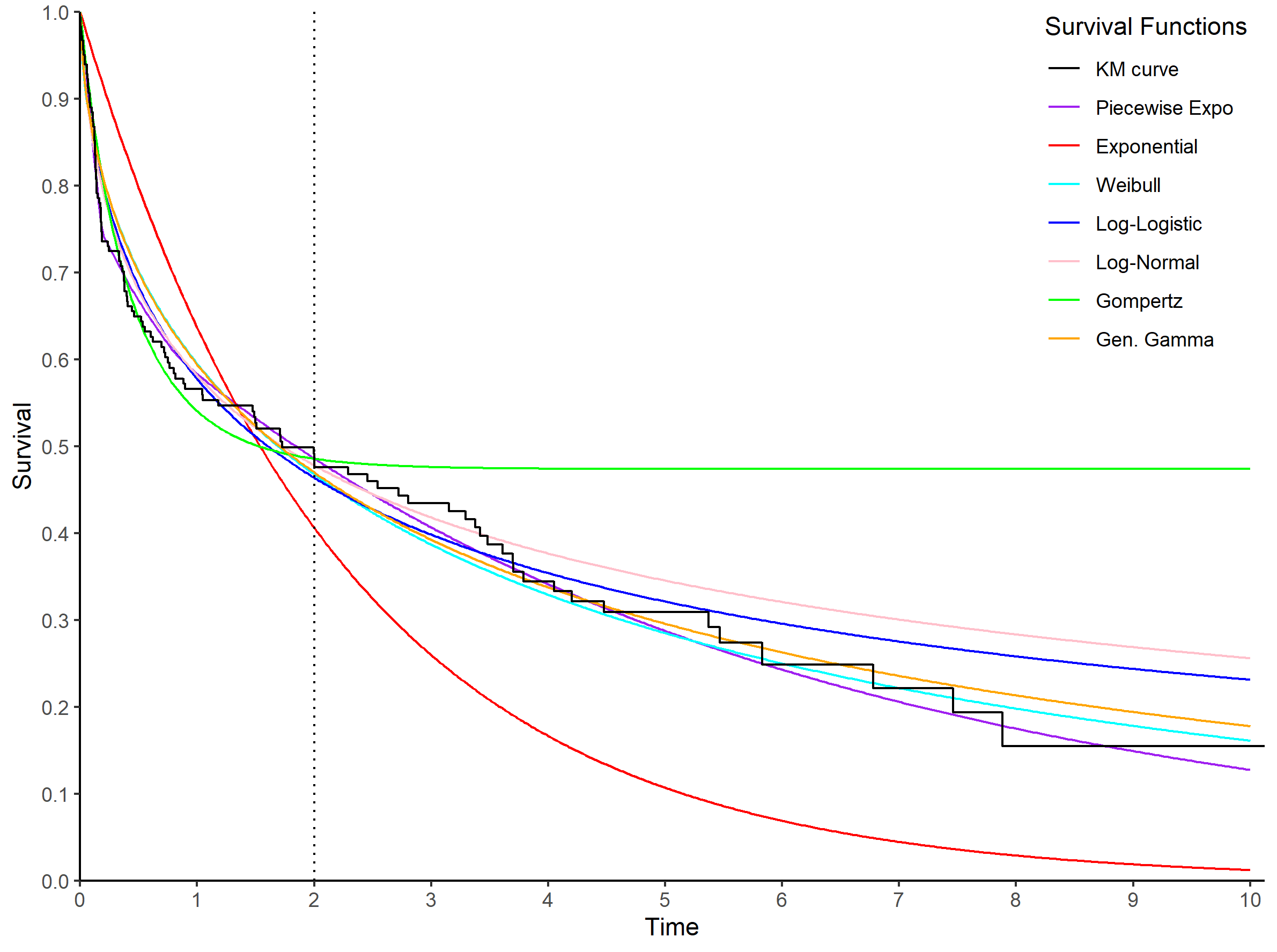} }
\caption{Long term survival probabilities for various models }
\label{fig:surv-RJMCMC}
\end{figure}

\begin{table}[h]
\centering
\caption{Predicted model fit to the long term data}
\begin{threeparttable}
 \begin{tabular}{c c c c c } 
  \hline
  Model						& -2log(PML)* & WAIC & AUC  & Absolute Difference  \\ 
  \hline
 Piecewise Exponential      & 248.45  & 248.26      & 3.35  & 0.14 \\
 Exponential               & 319.72   & 319.72       & 2.22  & 1.47 \\

 Weibull                   &  271.33   & 271.33     & 3.41  & 0.22  \\
 Log Logistic               & 262.11   & 262.11       & 3.71  & 0.43                \\ 
 Log Normal 				 & 265.50 & 265.51          & 3.91  & 0.55               \\ 
 Gompertz         & 263.22   & 263.22        & 4.99  & 1.63  \\
 Generalized Gamma         & 270.41   & 270.41        & 3.44  & 0.22  \\
  True Observations         &        &         & 3.42      & \\
 \hline
\end{tabular}
\begin{tablenotes}
      \small
       \item * -2log(PML) was calculated to place it on the same scale as WAIC. Lower values indicate better fit.
    \end{tablenotes}
\end{threeparttable}
\label{tab:output-collapsing}
\end{table}

\section{Discussion  \label{section Discussion}}

We have presented a Bayesian approach to determining the number and location of change-points in a hazard function including the special case were no change-point exists. By employing a Bayesian approach, uncertainty around the number of change-points is automatically computed and is described with a probabilistic interpretation, allowing us to assess the relative evidence of alternative change-point models. 

This approach takes advantage of the fact that the marginal likelihood for a piecewise exponential model without covariates can be expressed analytically. By restricting the change-points to be event times we reduce the complexity of the parameter space resulting in a simpler and computationally efficient algorithm. While the approach of \cite{Chapple.2020} is clearly more general in that it allows covariates and continuous change-points, we found that in some examples the change-points were highly correlated due to the relative infrequency of moves between model dimensions and that for smaller changes in the hazard change-points were not detected. Based on extensive testing on real world datasets and simulated examples, our approach demonstrated good mixing and rapid convergence throughout, attributes which are often difficult to achieve in problems in which the dimension of the parameter space can vary. This suggests that it is advantageous to adopt the collapsed approach introduced in this paper when covariates are not present. 

As with any Bayesian analysis, the inferences are somewhat informed by the choice of prior. We consider a discrete prior on the change-point locations which has the advantage of ensuring that the change-points are not too close together or close to the final events, where there is typically a sparsity of data. Because model selection is based on the evaluation of the marginal likelihood, this calculation can be sensitive to the hyperparameters used, and we provide an approach to specify a hyperprior for the $\beta$ hyperparameter which we describe in Web Appendix 1. The Poisson prior on the change-point number is reasonably robust to alternative specifications, however, in individual examples where posterior model probabilities are similar, it will naturally have some effect. We believe that a rate equal 1 for this Poisson prior is an appropriately parsimonious \emph{a priori} choice. 

Our simulation study demonstrates the ability of the algorithm to detect change-points when sample size and/or change in hazards is large along with the consistency of the estimators. As demonstrated by the no change-point simulation study, the model has a low probability of detecting the presence of change-points when they do not exist. 

In this paper we have presented two real world applications of the models. Regarding the Glioblastoma data, our approach segments the hazard function into distinct intervals which may allow greater interpretability of the trends in the hazard function even when not considering the piecewise exponential model for survival extrapolation. In situations where the constant long term hazards are plausible we believe that a piecewise exponential model should be considered. Although we artificially censored the data for the purpose of our example it is reasonable to hypothesize that these heart transplant patients may be subject to different hazards as time progresses. Patients are likely subject to high hazards of death during or immediately after a complex surgical procedure such as a heart transplant. Over the initial number of months patients are likely to be at an elevated risk of transplant rejection and many events may occur over this period. If patients do not reject their transplanted organ, over the long term they are subject to a lower hazard associated with all-cause mortality. With that point in mind piecewise exponential models (and any parametric model) should always be adjusted to ensure the extrapolated hazards do not fall below general population mortality. 

Finally we note the relatively recent Technical Support Document (TSD) regarding flexible survival models \cite{rutherford.2020}. Regarding the use piecewise exponential models in health technology assessment, they state that ``the cut-points for the various intervals may be arbitrary and may  importantly influence the results of an analysis'' and ``splitting the data into sections according to time means that sample sizes are reduced in later segments of the curve''. We believe our approach addresses both of these limitations as firstly the location (and number) of change-points is informed by the data and secondly the prior we use for the change-points reduces the probability that change-points close together or to the final event will be selected.
\cite{rutherford.2020} also highlight situations in which the Kaplan Meier survival function is used to represent the initial section of the survival function and an exponential function is adjoined to a predetermined point of the Kaplan Meier. In this situation, our approach could also be used in determining the final change-point and it's associated uncertainty from which the constant hazard is extrapolated.

\section{Declaration of Conflicting Interests}

Philip Cooney is an employee of Novartis Pharmaceutical Company and they have sponsored his PhD fees, however, Novartis has no input on the direction, output or content of the PhD. All statements within this document are the authors own opinions and do not represent the views of Novartis Pharmaceutical Company. 
Arthur White has no conflicts of interest to declare.

\backmatter


\bibliographystyle{biom}
\bibliography{ReferencesBiometrics}

\begin{thebibliography}{}

\bibitem[\protect\citeauthoryear{A.~E.~Gelfand}{A.~E.~Gelfand}{1994}]{Gelfand.1994}
A.~E.~Gelfand, D. K.~D. (1994).
\newblock Bayesian model choice: Asymptotics and exact calculations.
\newblock {\em Journal of the Royal Statistical Society. Series B
  (Methodological)} {\bf 56,} 501--514.

\bibitem[\protect\citeauthoryear{Anis}{Anis}{2009}]{Anis.2009}
Anis, M.~Z. (2009).
\newblock {Inference on a Sharp Jump in Hazard Rate: A Review}.
\newblock {\em Stochastics and Quality Control} {\bf 24,} 213--229.

\bibitem[\protect\citeauthoryear{Arjas and Gasbarra}{Arjas and
  Gasbarra}{1994}]{Arjas.1994}
Arjas, E. and Gasbarra, D. (1994).
\newblock Nonparametric bayesian inference from right censored survival data,
  using the gibbs sampler.
\newblock {\em Statistica Sinica} {\bf 4,} 505--524.

\bibitem[\protect\citeauthoryear{Bagust and Beale}{Bagust and
  Beale}{2014}]{Bagust.2014}
Bagust, A. and Beale, S. (2014).
\newblock Survival analysis and extrapolation modeling of time-to-event
  clinical trial data for economic evaluation: An alternative approach.
\newblock {\em Medical Decision Making} {\bf 34,} 343--351.
\newblock PMID: 23901052.

\bibitem[\protect\citeauthoryear{Castelloe and Zimmerman}{Castelloe and
  Zimmerman}{2002}]{Castelloe.2002}
Castelloe, J.~M. and Zimmerman, D.~L. (2002).
\newblock {Convergence Assessment for Reversible Jump MCMC Samplers}.
\newblock Technical report, University of Iowa.

\bibitem[\protect\citeauthoryear{Chapple, Peak, and Hemal}{Chapple
  et~al.}{2020}]{Chapple.2020}
Chapple, A., Peak, T., and Hemal, A. (2020).
\newblock A novel bayesian continuous piecewise linear log-hazard model, with
  estimation and inference via reversible jump markov chain monte carlo: Na.
\newblock {\em Statistics in Medicine} {\bf 39,}.

\bibitem[\protect\citeauthoryear{Davies, Briggs, Lorgelly, Garellick, and
  Malchau}{Davies et~al.}{2013}]{Davies.2013}
Davies, C., Briggs, A., Lorgelly, P., Garellick, G., and Malchau, H. (2013).
\newblock The “hazards” of extrapolating survival curves.
\newblock {\em Medical Decision Making} {\bf 33,} 369--380.
\newblock PMID: 23457025.

\bibitem[\protect\citeauthoryear{Fearnhead}{Fearnhead}{2006}]{Fearnhead.2006}
Fearnhead, P. (2006).
\newblock {Exact and efficient Bayesian inference for multiple changepoint
  problems}.
\newblock {\em Statistics and Computing} {\bf 16,} 203--213.

\bibitem[\protect\citeauthoryear{Goodman, Li, and Tiwari}{Goodman
  et~al.}{2011}]{Goodman.2011}
Goodman, M.~S., Li, Y., and Tiwari, R.~C. (2011).
\newblock {Detecting multiple change points in piecewise constant hazard
  functions}.
\newblock {\em Journal of applied statistics} {\bf 38,} 2523--2532.

\bibitem[\protect\citeauthoryear{Green}{Green}{1995}]{Green.1995}
Green, P.~J. (1995).
\newblock Reversible jump markov chain monte carlo computation and bayesian
  model determination.
\newblock {\em Biometrika} {\bf 82,} 711--732.

\bibitem[\protect\citeauthoryear{Hagar and Dukic}{Hagar and
  Dukic}{2015}]{Hagar.2015}
Hagar, Y. and Dukic, V. (2015).
\newblock {Comparison of hazard rate estimation in R}.

\bibitem[\protect\citeauthoryear{Han, Schell, and Kim}{Han
  et~al.}{2014}]{Han.2014}
Han, G., Schell, M.~J., and Kim, J. (2014).
\newblock {Improved survival modeling in cancer research using a reduced
  piecewise exponential approach}.
\newblock {\em Statistics in medicine} {\bf 33,} 59--73.

\bibitem[\protect\citeauthoryear{Hess and Gentleman}{Hess and
  Gentleman}{2019}]{muhaz.2019}
Hess, K. and Gentleman, R. (2019).
\newblock {\em muhaz: Hazard Function Estimation in Survival Analysis}.
\newblock R package version 1.2.6.1 (accessed Feb 01, 2021).

\bibitem[\protect\citeauthoryear{Incerti and Jansen}{Incerti and
  Jansen}{2020}]{hesim.2020}
Incerti, D. and Jansen, J.~P. (2020).
\newblock {\em {hesim: Health-Economic Simulation Modeling and Decision
  Analysis}}.
\newblock R package version 0.3.1 (accessed Feb 01, 2021).

\bibitem[\protect\citeauthoryear{Kearns, Stevenson, Triantafyllopoulos, and
  Manca}{Kearns et~al.}{2019}]{Kearns.2019}
Kearns, B., Stevenson, M.~D., Triantafyllopoulos, K., and Manca, A. (2019).
\newblock Generalized linear models for flexible parametric modeling of the
  hazard function.
\newblock {\em Medical Decision Making} {\bf 39,} 867--878.
\newblock PMID: 31556792.

\bibitem[\protect\citeauthoryear{Kim, Cheon, and Jin}{Kim
  et~al.}{2020}]{Kim.2020}
Kim, J., Cheon, S., and Jin, Z. (2020).
\newblock {Bayesian multiple change-points estimation for hazard with censored
  survival data from exponential distributions}.
\newblock {\em Journal of the Korean Statistical Society} {\bf 49,} 15--31.

\bibitem[\protect\citeauthoryear{Kosinski and Biecek}{Kosinski and
  Biecek}{2020}]{RTCGA}
Kosinski, M. and Biecek, P. (2020).
\newblock {\em RTCGA: The Cancer Genome Atlas Data Integration}.
\newblock R package version 1.18.0 (accessed Feb 01, 2021).

\bibitem[\protect\citeauthoryear{Loader}{Loader}{1991}]{Loader.1991}
Loader, C.~R. (1991).
\newblock {Inference for a Hazard Rate Change Point}.
\newblock {\em Biometrika} {\bf 78,} 749--757.

\bibitem[\protect\citeauthoryear{Matthews and Farewell}{Matthews and
  Farewell}{1982}]{Matthews.1982}
Matthews, D.~E. and Farewell, V.~T. (1982).
\newblock {On Testing for a Constant Hazard against a Change-Point
  Alternative}.
\newblock {\em Biometrics} {\bf 38,} 463--468.

\bibitem[\protect\citeauthoryear{Miller and Halpern}{Miller and
  Halpern}{1982}]{Miller.1982}
Miller, R. and Halpern, J. (1982).
\newblock {Regression with Censored Data}.
\newblock {\em Biometrika} {\bf 69,} 521--531.

\bibitem[\protect\citeauthoryear{Müller and Wang}{Müller and
  Wang}{1994}]{Muller.1994}
Müller, H.~G. and Wang, J.-L. (1994).
\newblock {Change-Point Models for Hazard Functions}.
\newblock {\em Lecture Notes-Monograph Series} {\bf 23,} 224--241.

\bibitem[\protect\citeauthoryear{Paola~Rebora and Reilly}{Paola~Rebora and
  Reilly}{2018}]{bshazard.2018}
Paola~Rebora, A.~S. and Reilly, M. (2018).
\newblock {\em {bshazard: Nonparametric Smoothing of the Hazard Function}}.
\newblock R package version 1.1 (accessed Feb 01, 2021).

\bibitem[\protect\citeauthoryear{Plummer}{Plummer}{2003}]{Plummer.2003}
Plummer, M. (2003).
\newblock Jags: A program for analysis of bayesian graphical models using gibbs
  sampling.

\bibitem[\protect\citeauthoryear{Raftery}{Raftery}{1986}]{Raftery.1986}
Raftery, A. (1986).
\newblock Choosing models for cross-classifications.
\newblock {\em American Sociological Review} {\bf 51,} 145.

\bibitem[\protect\citeauthoryear{Rutherford, Lambert, Sweeting, Pennington,
  Crowther, Abrams, and Latimer}{Rutherford et~al.}{2020}]{rutherford.2020}
Rutherford, M.~J., Lambert, P.~C., Sweeting, M.~J., Pennington, B., Crowther,
  M.~J., Abrams, K.~R., and Latimer, N.~R. (2020).
\newblock Flexible methods for survival analysis tsd.

\bibitem[\protect\citeauthoryear{Vehtari, Gabry, Magnusson, Yao, Bürkner,
  Paananen, and Gelman}{Vehtari et~al.}{2020}]{loo.2020}
Vehtari, A., Gabry, J., Magnusson, M., Yao, Y., Bürkner, P.-C., Paananen, T.,
  and Gelman, A. (2020).
\newblock loo: Efficient leave-one-out cross-validation and waic for bayesian
  models.
\newblock R package version 2.4.1.

\bibitem[\protect\citeauthoryear{Wang, Dignam, Won, Curran, Mehta, and
  Gilbert}{Wang et~al.}{2015}]{Wang.2015}
Wang, M., Dignam, J.~J., Won, M., Curran, W., Mehta, M., and Gilbert, M.~R.
  (2015).
\newblock {{Variation over time and interdependence between disease progression
  and death among patients with glioblastoma on RTOG 0525}}.
\newblock {\em Neuro-Oncology} {\bf 17,} 999--1006.

\bibitem[\protect\citeauthoryear{Watanabe}{Watanabe}{2010}]{Watanabe.2010}
Watanabe, S. (2010).
\newblock Asymptotic equivalence of bayes cross validation and widely
  applicable information criterion in singular learning theory.
\newblock {\em J. Mach. Learn. Res.} {\bf 11,} 3571–3594.

\bibitem[\protect\citeauthoryear{Wyse and Friel}{Wyse and
  Friel}{2010}]{Wyse.2010}
Wyse, J. and Friel, N. (2010).
\newblock {Simulation-based Bayesian analysis for multiple changepoints}.

\bibitem[\protect\citeauthoryear{Yao}{Yao}{1986}]{Yao.1986}
Yao, Y.-C. (1986).
\newblock {Maximum likelihood estimation in hazard rate models with a
  change-point}.
\newblock {\em Communications in Statistics - Theory and Methods} {\bf 15,}
  2455--2466.

\end{thebibliography}

\section*{Supporting Information}
Web Appendix A, referenced in Section~\ref{section Discussion}, is available with
this paper at the Medical Decision Making website. Method implementation in a R package, as well as a worked example using a simulated dataset can be found at https://github.com/Philip-Cooney/PiecewiseChangepoint.
\vspace*{-8pt}

\subsection*{Incorporating uncertainty in hyperparameters}
\label{section:uncertain_hyperparameters}

The marginal likelihood can be sensitive to the hyperparameters $\alpha$ and $\beta$ and therefore can influence the posterior distribution of the change-points. To account for this uncertainty and improve the robustness of the results we can introduce a hyperprior on $\beta$. In an extra sampling step, the hazards $\lambda_{1:k+1}$ can be ``uncollapsed'' and sampled at each iteration. 
We place a hyperprior on $\beta:$ $\beta \sim \mathcal G(\xi, \delta)$, with
\[ \pi(\beta | \xi, \delta) = \frac{\delta^\xi}{\Gamma(\xi)} \beta^{\xi -1} \exp\left(-\delta \beta \right). \]

\noindent
Simplifying Equation \ref{eq:full_posterior} we note that the posterior density of the change-point number and locations is proportional to the likelihood, the prior on the hazards and the hyperprior on $\beta$. 

\begin{eqnarray*}
 \pi(k,s_1,\dots,s_k,\beta|y_{1:d},\lambda_{1:k+1},  \alpha,\xi, \delta) &\propto & \pi \left
 (y_{1:d}|s_1,\dots,s_k,\lambda_{1:k+1} \right)\prod ^{k+1}_{j=1}\pi(\lambda_j | \alpha, \beta) \times \pi(\beta | \xi, \delta) \\
&= &   \prod_{j=1}^{k+1} \Bigg[ \lambda_j^{(s_j - s_{j-1})} - \text{exp}^{\lambda_j\sum_{i=s_{(j-1) +1}}^{s_j} y_i}\Bigg] \\ &\times& \prod ^{k+1}_{j=1} \frac{\beta^\alpha}{\Gamma(\alpha)} \lambda_j^{\alpha -1} \exp\left(\beta\lambda_j \right) \times \frac{\delta^\xi}{\Gamma(\xi)} \beta^{\xi -1} \exp\left(\delta \beta \right) .
\end{eqnarray*}

\noindent
The marginal distribution of $\pi ( \beta | k,s_1,\dots,s_k,y_{1:d},\lambda_{1:k+1},  \alpha,\xi, \delta)$ is 

\begin{eqnarray*}
\pi (\beta |k,s_1,\dots,s_k,y_{1:d},\lambda_{1:k+1},\alpha,\xi, \delta) &\propto & \prod_{j=1}^{k+1} \Bigg[ \lambda_j^{(s_j - s_{j-1})} - \text{exp}^{\lambda_j\sum_{i=s_{(j-1)}+1}^{s_j} y_i}\Bigg] \\ &\times& \prod ^{k+1}_{j=1} \frac{\beta^\alpha}{\Gamma(\alpha)} \lambda_j^{\alpha -1} \exp\left(\beta\lambda_j \right) \times \frac{\delta^\xi}{\Gamma(\xi)} \beta^{\xi -1} \exp\left(-\delta \beta \right) \\
&\propto &  \prod ^{k+1}_{j=1} \beta^\alpha \exp\left(\beta\lambda_j \right) \times \beta^{\xi -1} \exp\left(-\delta \beta \right)  \\
&=&  \beta^{(k + 1)\alpha + \xi -1} \exp\left(-\beta \Bigg[\sum ^{k+1}_{j=1}\lambda_j + \delta \Bigg] \right).
\end{eqnarray*}

This is the kernel of a gamma distribution with shape $(k + 1)\alpha + \xi$ and rate $\sum ^{k+1}_{j=1}\lambda_j + \delta$ and is updated once each iteration. The hazards are sampled from a gamma distribution $\lambda_{j}|\alpha, \beta, k~\sim~\mathcal{G}(\alpha + s_j-s_{j-1}, \beta^* + \sum_{i=s_{(j-1) +1}}^{s_j} y_i)$ with $\beta^*$ the current value of $\beta$ before the sampling of a new $\beta$. 

One important practical point relates to the choice of $\xi, \delta$ when the data has a particular timescale. If the data is in years we should set $\beta$ to have an expected value of 1 (assuming that in all cases $\alpha$ is set to 1) with variance 1 which is achieved by setting $\xi =1, \delta=1$ and follows immediately from the properties of the Gamma distribution. If the data is in days and we wish to retain the same prior we require $\beta$ to have an expected value and variance of 365 which is achieved $\xi =1, \delta=1/365$.

\appendix

\section{}
\subsection{Marginal Likelihood for exponential survival times}
\label{section:marg-like-deriv}
The marginal likelihood is sometimes known as the probability of the data and appears as the denominator in Bayes Formula 

$$ \pi(\mathbf{y}) = \int_{\theta} \pi(\mathbf{y}|\theta)\pi(\theta)d\theta.$$
\noindent
In our model we consider the data conditional on the hyperparameters $\alpha$ and $\beta$ which we denote together as $\gamma$. Therefore the expression becomes

$$ \pi(\mathbf{y}|\gamma) = \int_{\theta} \pi(\mathbf{y}|\theta)\pi(\theta|\gamma)d\theta.$$

\noindent
The easiest way to evaluate this integral is indirectly through Bayes formula. Bayes formula is as follows:
$$ \pi(\theta|\mathbf{y}, \gamma) = \frac{\pi(\mathbf{y}|\theta)\pi{(\theta)}\pi(\gamma)}{\int_{\theta} \pi(\mathbf{y}|\theta)\pi(\theta|\gamma)d\theta}.$$ 
The conjugate prior for an exponential likelihood is the gamma distribution. Therefore given hyperparameters $\alpha$ and $\beta$, the posterior is a $\mathcal{G}(\alpha + D , \beta + T)$ where $D$ is the number of events and $T$ is the exposure time within that interval. Letting $\alpha^* = \alpha + D$ and  $\beta^* = \beta + T$ and rearranging Bayes formula, it immediately follows that the marginal likelihood is the ratio of the prior normalizing factor divided by the posterior normalizing factor; 

$$ \int_{\theta} \pi(\mathbf{y}|\theta)\pi(\theta|\gamma)d\theta = \frac{\pi(\mathbf{y}|\theta)\pi{(\theta)}\pi(\gamma)}{\pi(\theta|\mathbf{y})}= \frac{\beta^\alpha/\Gamma(\alpha)}{({\beta^*}^{\alpha^*})/\Gamma(\alpha^*)}.$$

\noindent
Given $k$ change-points, we have $k+1$ segments of data and the joint marginal likelihood is the product of these $k+1$ segments.

\label{lastpage}
\end{document}